\documentclass[final]{article}

\usepackage[dblblindworkshop, nonatbib]{neurips_2026}
\workshoptitle{Interpretability for Discovery}

\usepackage[utf8]{inputenc} 
\usepackage[T1]{fontenc}    
\usepackage{hyperref}       
\usepackage{url}            
\usepackage{booktabs}       
\usepackage{amsfonts}       
\usepackage{amsmath}        
\usepackage{amssymb}        
\usepackage{microtype}      
\usepackage{xcolor}         
\usepackage{graphicx}       
\usepackage{rotating}       
\usepackage{sidecap}
\usepackage{subcaption}
\usepackage[
    backend=biber,
    natbib=true,
    style=authoryear-comp,
]{biblatex}

\title{Gestalt: a meta-foundation model for astronomy}

\author{%
\begin{tabular}{c}
Michael~J.~Smith$^{1,2,3}$ \qquad Shashwat Sourav$^{3,4}$\\
\normalfont \texttt{mike@mjjsmith.com} \qquad \texttt{s.shashwat@wustl.edu}\\[8pt]
\normalfont $^{1}$AstroAI,
\normalfont $^{2}$Center for Astrophysics $\vert$ Harvard \& Smithsonian, \\
\normalfont $^{3}$UniverseTBD, $^{4}$Washington University in St. Louis
\end{tabular}
}

\begin{document}

\maketitle

\begin{abstract}
The Platonic Representation Hypothesis predicts that sufficiently scaled foundation models converge on a shared representation of the world.
As each non-converged model gives a noisy view of a common structure when passed the same input, we ask whether we can combine models into a representation that outperforms its individual components.
We test this on galaxies: we embed images via a basket of 22 frozen foundation models from eight families, whiten each view, and take a randomised SVD of the embedding concatenation.
The resulting 1024-dimensional embedding outperforms every basket member on 19/21 of our tested metrics for physical property and galaxy morphology estimation for HSC, JWST, and DESI Legacy Survey imagery.
We find that performance rises with basket size and basket architectural diversity, and that the meta-foundation model's performance transfers across astronomical surveys.
    We conclude that a useful astronomical foundation model can be assembled from existing generalist models with no training required beyond a single unsupervised projection.
    By leveraging the community's already-spent work, we save a lot of compute: a fresh pre-train of a comparable single-domain model would cost $\mathcal{O}(10^{4}$--$10^{5})$ A100 GPU hours (emitting several tonnes of CO$_2$eq.), whereas assembling Gestalt requires minutes on a single machine.
\end{abstract}

\section{The fourth wave of astro-connectionism, and a question}
\label{sec:intro}

Four waves of connectionism have lapped astronomy's shores \citep{ref_smith2023}. 
The first brought small MLPs tuned on hand-picked features \citep[e.g.][]{ref_adorf1988,ref_angel1990}; the second swept those away with CNNs and RNNs trained end-to-end on raw data \citep[e.g.][]{ref_dieleman2015,ref_charnock2018}; the third introduced self- and unsupervised representation learning \citep[e.g.][]{ref_smith2021,ref_sarmiento2021}, removing the need for human-defined labelling.
The fourth, now breaking, has gifted astronomy an embarrassment of \emph{foundation models}: large, expensively pre-trained networks whose internal representations transfer linearly to a remarkable range of scientific tasks. Some are
trained on natural images and applied few- or zero-shot to galaxies \citep{ref_lastufka2024,ref_universetbd2025}; some are trained on galaxy data directly \citep{ref_smith2024,ref_parker2024}; some are trained on text \citep{ref_nguyen2023,ref_heneka2026}, audio \citep{ref_chatterjee2024}, or none of the above \citep{ref_hirashima2025,ref_vega2025}. 
Only the specialists can be said to have any privileged claim to astronomical knowledge, and yet most of them, presented with a galaxy, will return an embedding from which useful information can be read off with a linear or MLP probe \citep{ref_universetbd2025}. 

The natural engineering response is to pick the best single model and get on with the science, but we believe that the Platonic Representation Hypothesis (PRH) \citep{ref_huh2024} implies the existence of an alternative approach.
The PRH proposes that sufficiently large networks trained on diverse data converge toward a shared statistical model of reality, regardless of architecture and objective, with direct evidence for this convergence across astronomical surveys recently reported \citep{ref_universetbd2025}. 
If a set of frozen networks project the same galaxy through a set of correlated transforms, we ask whether the physical information carried by the galaxy can be recovered more faithfully from some summary embedding space than through any single model.
We find that this is the case through a simple self-supervised process of denoising, rotation, concatentation, and projection of the models' embeddings.

We present the following key results: (1) Our meta-model beats every model in our basket for 19/21 of our tested physical and morphological metrics.
(2) Performance scales with our tested model basket size and diversity.
(3) The physical probe geometry is reorganised beyond the full basket range, recovering the pattern predicted by astrophysics.
(4) Performance transfers to held-out surveys when trained on a multi-survey dataset.
(5) The artifact producing the embedding is $\mathcal{O}(100)$\,MB and applies to new data as a single matrix multiplication; we release the framework as \emph{Gestalt}.

\begin{figure}
    \centering
    \includegraphics[width=\linewidth]{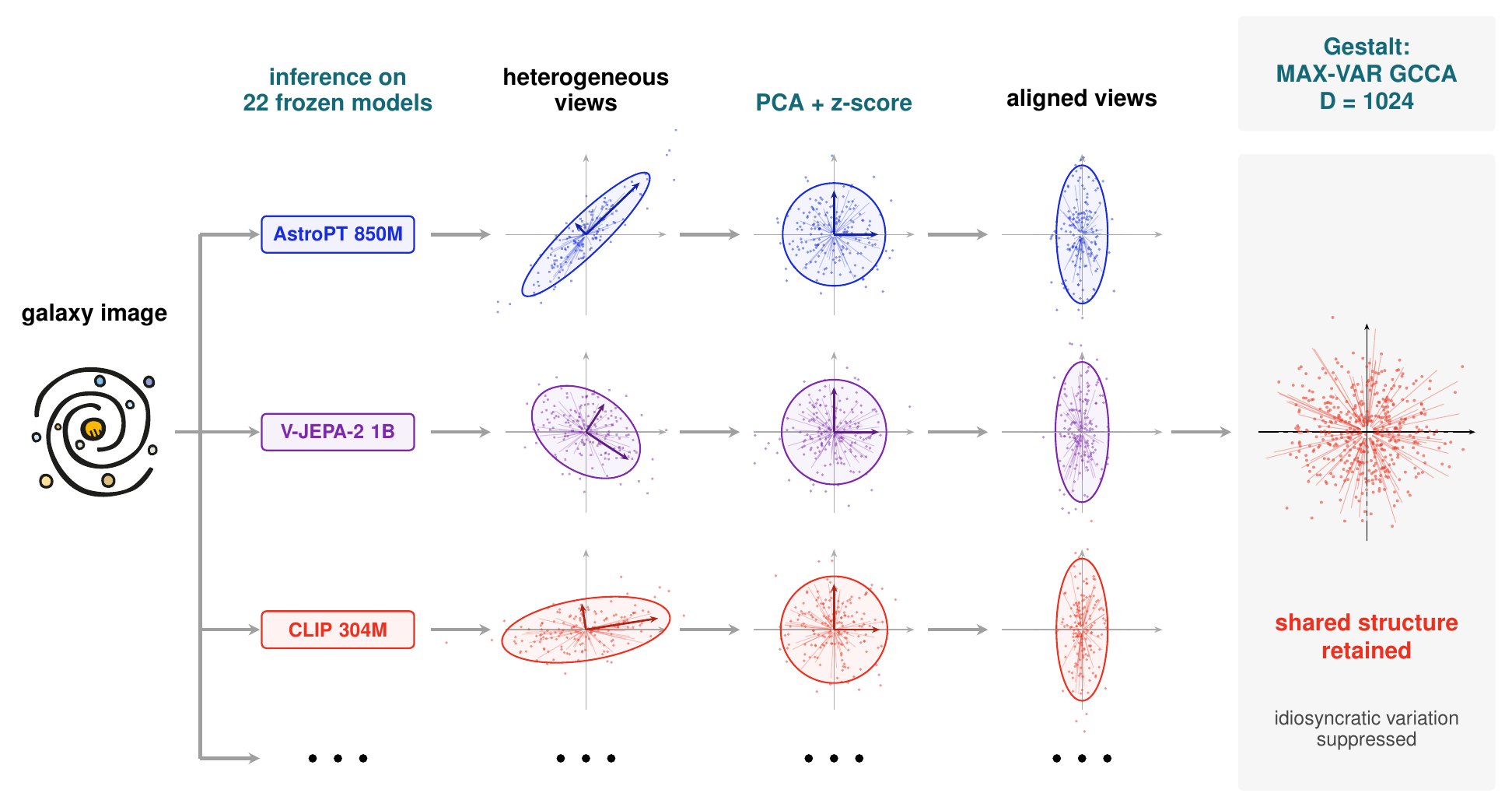}
    \caption{The Gestalt meta-model.}
    \label{fig:gestalt}
\end{figure}

\section{The Gestalt meta-model}
\label{sec:method}

We view each frozen foundation model as a fixed encoder $E^{(m)} : X \to \mathbb{R}^{d_m}$ from galaxy cutouts to its own representation space. 
If the PRH holds, there is a subspace common to all $M$ encoders on which they encode the same physical content, up to a linear transform of each. 
We aim to identify it without further pre-training through the following procedure (see also Fig.~\ref{fig:gestalt}):

We assemble $M=22$ publicly available frozen image foundation-model checkpoints from eight architecture families (Tab.~\ref{tab_basket}), spanning pooled-output widths $d_m$ from 384 to 5120 and a deliberately wide range of training paradigms from supervised classification \citep{ref_dosovitskiy2020,ref_woo2023}, to masked-image reconstruction \citep{ref_he2022}, to joint-embedding prediction \citep{ref_assran2023,ref_assran2025}, to contrastive image--text alignment \citep{ref_radford2021}, to visual instruction tuning\citep{ref_liu2023llava2}, and one in-domain autoregressive transformer \citep{ref_smith2024}. 

\begin{SCtable}[0.6][ht]
  \caption{Our tested basket containing 22 frozen checkpoints across eight families. 
  `$d_m$' is the model's pooled-output width; counts and widths correspond left-to-right within each row.}
  \label{tab_basket}
  \centering\tiny
  \begin{tabular}{lll}
    \toprule
    Family & Parameter counts & Pooled-output widths $d_m$ \\
    \midrule
    AstroPT \citep{ref_smith2024}       & 15M, 95M, 850M        & 384, 768, 2048 \\
    I-JEPA \citep{ref_assran2023}       & 632M, 1B              & 1280, 1408 \\
    V-JEPA-2 \citep{ref_assran2025}     & 300M, 600M, 1B        & 1024, 1280, 1408 \\
    ViT \citep{ref_dosovitskiy2020}     & 86M, 304M, 632M       & 768, 1024, 1280 \\
    ViT-MAE \citep{ref_he2022}          & 86M, 304M, 632M       & 768, 1024, 1280 \\
    CLIP \citep{ref_radford2021}        & 86M, 304M             & 512, 768 \\
    LLaVA-1.5 \citep{ref_liu2023llava2} & 7B, 13B               & 4096, 5120 \\
    ConvNeXt-V2 \citep{ref_woo2023}     & 15M, 28M, 89M, 198M  & 640, 768, 1024, 1536 \\
    \bottomrule
  \end{tabular}
\end{SCtable}

To prevent any view's native scale from dominating the joint SVD, we apply a full-rank PCA separately to each model embedding and z-score every principal component on the fit sample.

We concatenate the native-width whitened views into \(C\in\mathbb{R}^{N\times\sum_m d_m}\) and run a rank-\(D\) randomized singular value decomposition (SVD) with $D=1024$ \citep{ref_halko2011}: $C \approx U\Sigma V^{\!\top},\,\text{Gestalt} \equiv CV \approx U\Sigma.$ This is equivalent to a MAX-VAR GCCA \citep{ref_carroll1968,ref_kettenring1971} where $U$ captures the directions of greatest agreement across views, and $\text{Gestalt} \in \mathbb{R}^{N \times D}$ gives the corresponding joint embedding. 
New data are embedded by applying the whitening, concatenating the transformed views, and multiplying by $V$.
Under a PRH interpretation, these directions capture shared physical and instrumental content while discarding model-specific noise. 

We evaluate embeddings with the probe protocol of \citet{ref_universetbd2025}.
We standardise features, clip targets at the 1st and 99th percentiles, and run a linear regression scored by held-out $R^2$.
We run a randomized train/test split as follows: COSMOS-Web uses 40\,000/5\,000, GZ10 uses 15\,300/2\,500, and \texttt{Smith42/galaxies} holds out 5\,000 valid labels per target except tight spiral due to its small corpus size (where the test set is 735 of 3\,676 valid objects).

\section{Experimental results and discussion}
\label{sec:experiments}

We test Gestalt by estimating physical and morphological galaxy properties on three imaging datasets: Hyper Suprime-Cam (HSC) \citep{ref_miyazaki2018} and James Webb Space Telescope (JWST) \citep{ref_gardner2006} imagery provided by the COSMOS-Web catalog \citep{ref_casey2023}, and DESI Legacy Survey imagery \citep{ref_dey2019} provided by the catalogs \texttt{Smith42/galaxies} (v2.0) \citep{ref_smith2024,ref_sanjaripour2026} and Galaxy Zoo 10 (GZ10) \citep{ref_walmsley2022decals,ref_mmu2024}.

\textbf{Imagery preprocessing.}
Our encoders receive an identical RGB rendering of each galaxy per survey. 
Our COSMOS-Web cutouts are drawn from a crossmatched HSC/JWST catalogue\footnote{\url{https://hf.co/datasets/Ashodkh/cosmosweb-hsc-jwst-high-snr-pil2}}.
For HSC we use the $g$, $r$, and $z$ bands sampled onto a $96\times96$ pixel canvas.
For JWST we use the F090W, F277W, and F444W bands at their native size.
Every retained band is arcsinh-stretched, $\varphi=\operatorname{arcsinh}(20\,\phi)/\operatorname{arcsinh}(20)$, where $\phi$ is the band flux linearly rescaled by fixed per-survey normalisation constants (global 1st and 99th flux percentiles) and clipped to $[0,1]$. 
Our two remaining corpora are used as-is: both the DESI Legacy Survey cutouts of GZ10 and \texttt{Smith42/galaxies} (v2.0) pass through unchanged apart from bilinear resampling onto a $96\times96$ canvas.

\textbf{Gestalt beats the basket in 19/21 tested metrics.}
We fit Gestalt per imaging modality at $D=1024$ and find that Gestalt has the highest score in all six probed COSMOS-Web cases (photo-$z$, $\log M_\star$, and sSFR; Tab.~\ref{tab:probes}).
We also run the method described in \S\ref{sec:method} on a 10-class morphology classification plus redshift regression benchmark on GZ10, and the 13-target regression suite of the \texttt{Smith42/galaxies} v2.0 dataset, which contains absolute magnitudes, colours, spec-$z$, photo-$z$, sSFR, $\log M_\star$, and morphology vote fractions.
We find that Gestalt is the best performing model for GZ10 redshift and 12 of the 13 \texttt{Smith42/galaxies} targets, while native-width CLIP leads GZ10 morphology with Gestalt within error bounds (Tab.~\ref{tab:probes}).

\begin{table}[htbp]
\caption{
  Linear-probe scores (mean $\pm$ standard deviation across five random train--test splits); bold marks the highest mean in each row.
  For \texttt{Smith42/galaxies}, each split is summarized by the median over 13 targets, with the best single using the fixed model with the highest value of these split-wise medians.
  Full results are in the appendix.\\}
 \label{tab:probes}
 \centering\scriptsize
 \resizebox{\columnwidth}{!}{%
 \begin{tabular}{lllccc}
   \toprule
   dataset & target & metric & \textbf{Gestalt} & concat$\to$PCA & best single \\
   \midrule
   HSC COSMOS-Web  & $z$            & $R^2$ & \textbf{0.518}$\pm$.008 & 0.482$\pm$.008 & 0.453$\pm$.008 (AstroPT 850M) \\
   HSC COSMOS-Web & $\log M_\star$ & $R^2$ & \textbf{0.604}$\pm$.012 & 0.577$\pm$.009 & 0.560$\pm$.013 (V-JEPA-2 1B) \\
   HSC COSMOS-Web & sSFR           & $R^2$ & \textbf{0.581}$\pm$.015 & 0.555$\pm$.007 & 0.530$\pm$.008 (V-JEPA-2 1B) \\
   \midrule
   JWST COSMOS-Web & $z$            & $R^2$ & \textbf{0.658}$\pm$.014 & 0.595$\pm$.013 & 0.565$\pm$.012 (V-JEPA-2 1B) \\
   JWST COSMOS-Web & $\log M_\star$ & $R^2$ & \textbf{0.848}$\pm$.003 & 0.815$\pm$.004 & 0.810$\pm$.004 (V-JEPA-2 1B) \\
   JWST COSMOS-Web & sSFR           & $R^2$ & \textbf{0.482}$\pm$.007 & 0.458$\pm$.004 & 0.457$\pm$.010 (V-JEPA-2 1B) \\
   \midrule
   Galaxy Zoo 10     & morphology & $F_1$ & 0.708$\pm$.009 & 0.674$\pm$.009 & \textbf{0.713}$\pm$.011 (CLIP 304M) \\
   Galaxy Zoo 10     & $z$        & $R^2$ & \textbf{0.758}$\pm$.008 & 0.729$\pm$.010 & 0.688$\pm$.013 (I-JEPA 632M) \\
   \midrule
   \texttt{Smith42/galaxies} & 13 targets \citep{ref_sanjaripour2026} & $R^2$ & \textbf{0.779}$\pm$.005 & 0.747$\pm$.008 & 0.700$\pm$.002 (CLIP 304M) \\
   \bottomrule
 \end{tabular}%
 }
\end{table}

\begin{figure}[htbp]
  \begin{subfigure}{\linewidth}
    \centering
    \includegraphics[width=.82\linewidth]{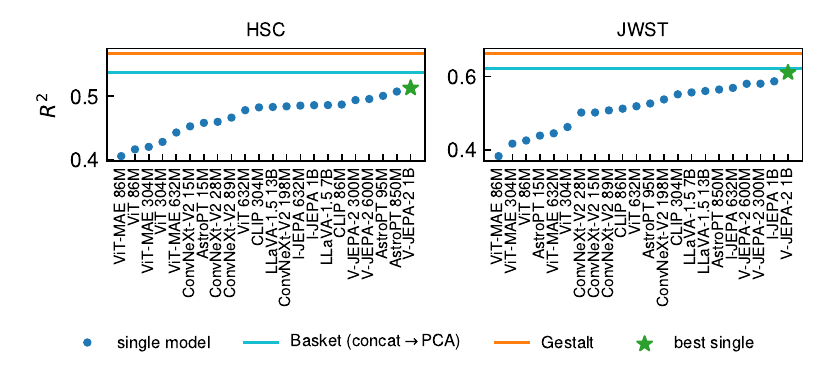}
    \caption{Mean linear-probe $R^2$ across five repeated random splits for the 22 basket members (blue), the best single member per modality (green star), Gestalt (orange line), and the unaligned concat$\to$PCA model (teal line).}
    \label{fig:cosmos-summary}
  \end{subfigure}
  \begin{subfigure}{\linewidth}
    \centering
    \includegraphics[width=.82\linewidth]{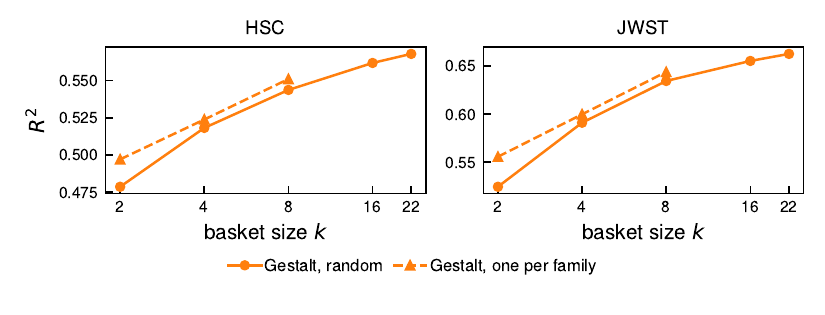}
    \caption{Mean linear-probe $R^2$ for the COSMOS-Web dataset vs.\ Gestalt model basket size $k$. The dashed line shows one-model-per-family subsets through $k=8$, the number of families; the solid line shows random subsets.}
    \label{fig:scaling}
  \end{subfigure}
  \begin{subfigure}{\linewidth}
    \centering
    \includegraphics[width=.82\linewidth]{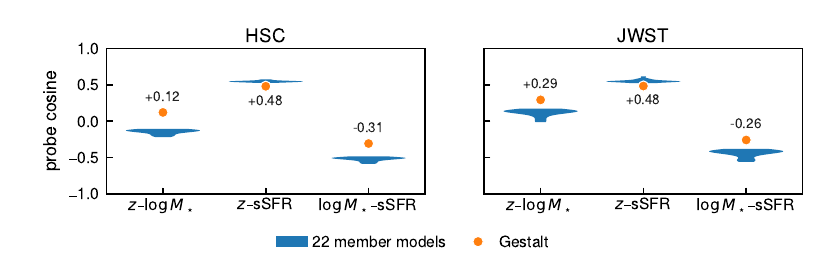}
    \caption{Violin plots show the distributions across 22 frozen foundation models of the pairwise cosine similarities between linear-probe directions for redshift ($z$), stellar mass ($M_\star$), and specific star-formation rate (sSFR); orange points show the corresponding values in the Gestalt representation.}
    \label{fig:probes}
  \end{subfigure}
  \begin{subfigure}{\linewidth}
    \centering
    \includegraphics[width=.8\linewidth]{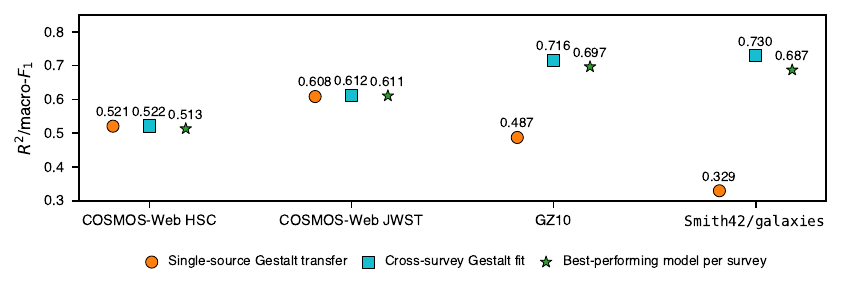}
      \caption{Mean downstream task score on each target corpus all tasks. 
      Orange circles average the `transfer score' from three Gestalt representations, each fitted on unlabelled objects from a single, different corpus to the tested corpus. 
      Blue squares show one Gestalt representation fitted on a merged dataset of 2\,500 unlabelled objects from each of the three other corpora. 
      Green stars show the best native single model for each corpus, shown here as a reference.}
    \label{fig:native-lodo-single}
  \end{subfigure}
  \caption{Experimental results investigating model performance (a), basket size scaling (b), physics subspace composition (c), and cross-survey transfer (d).}
\end{figure}

\phantomsection\label{sec:scaling}
\textbf{Performance scales with basket size and diversity.} When we refit Gestalt on random sub-baskets of size $k \in \{2, 4, 8, 16, 22\}$ (Fig.~\ref{fig:scaling}), we find that mean linear-probe $R^2$ across the six COSMOS-Web metrics rises monotonically from 0.501 at $k=2$ to 0.555, 0.589, 0.609, and 0.615 at $k=4,8,16,$ and 22, respectively. 
Choosing one model per architecture family beats random subsets at every matched $k\in\{2,4,8\}$, suggesting that architectural diversity is an important factor in realizing performance.

\textbf{Gestalt reorganises the physical geometry of its constituent representations.}                    
We now take the fitted COSMOS-Web Gestalt linear probes for redshift ($z$), stellar mass ($M_\star$), and specific star-formation rate (sSFR), and use the cosine similarity between their weight vectors to measure how these properties are organised within each representation.
A positive cosine means that increases in two properties are encoded along similar directions, whereas a negative cosine means that their probe directions oppose one another.
Figure~\ref{fig:probes} compares these cosines in Gestalt with their distributions across our models.
Astrophysically, we expect a sign pattern of $(+,+,-)$ in Fig.~\ref{fig:probes}:
\textbf{first}, the positive $z$--$M_\star$ direction primarily reflects Malmquist-style selection as only the more luminous and generally more massive galaxies remain detectable at high redshift in a flux-limited sample \citep{ref_malmquist1922};
\textbf{second}, the positive $z$--sSFR direction reflects the cosmic evolution of star formation: galaxies at higher redshift typically form stars more rapidly relative to their existing stellar mass \citep{2014ARA&A..52..415M};
\textbf{third}, the negative $M_\star$--sSFR direction follows from the galaxy main sequence, $\log\mathrm{SFR}=\alpha\log M_\star+\beta(z)$, whose slope is typically sub-linear ($\alpha<1$).
Since $\log\mathrm{sSFR}=\log\mathrm{SFR}-\log M_\star$, this gives $\log\mathrm{sSFR}=(\alpha-1)\log M_\star+\beta(z)$ and hence a negative dependence on mass \citep{2014ApJS..214...15S,ref_wetzel2012}.

Gestalt recovers this $(+,+,-)$ organisation in both HSC and JWST, while lying outside the complete member-model range for every property pair. 
We note that every HSC member gives a negative $z$--$M_\star$ cosine, whereas Gestalt changes its sign to positive.
Because Gestalt is fitted independently of the physical probes, these displacements cannot arise from target-specific optimisation.
They show that Gestalt is not an average of its members, and suggest that alignment constructs a new shared coordinate system in which their common information supports a distinct and more cross-modally consistent organization of galaxy properties.                    

\textbf{Gestalt generalizes across surveys.}
We find that Gestalt is able to generalize across surveys by fitting on three out of our four tested surveys and then training linear probes on the remainder.
Figure~\ref{fig:native-lodo-single} shows perfomance as defined as the average over that corpus' downstream tasks.
For each target corpus the `single-source Gestalt transfer' takes the Gestalt models fitted individually on each of the three \emph{other} corpora and applies them unchanged to the target, taking the mean score across models for Fig.~\ref{fig:native-lodo-single}.
The cross-survey Gestalt fit takes a projection on unlabelled objects pooled from the three other corpora, with no target-survey imagery entering the fit (for the two COSMOS-Web corpora the pool does contain the companion telescope's views of the same galaxies; see App.~\ref{app:cross-survey}).
Finally, the best-performing model per survey is the strongest individual frozen model evaluated natively on its own survey at $D=1024$.
We provide this as an upper reference point rather than a protocol-matched comparison to the transfer fits.
We find that Gestalt performs well across surveys when its training set consists of cross-survey imagery, even if the specific tested survey imagery is held out.

\textbf{Limitations.} 
We evaluate via linear probing to keep comparisons between models simple, and leave more complicated probing methods to future work.
Embedding a new galaxy requires a forward pass through all $M$ encoders, so deployment carries roughly $M\times$ the inference cost of any single basket member.
Gestalt therefore trades a one-off pre-training expense for a recurring per-galaxy inference overhead.
Future work will include attempts to find a way to run more efficient inference on Gestalt.
Our basket is derived from `well-behaved' models taken from \citet{ref_universetbd2025}, and we expect performance to change with basket composition; due to compute limitations we leave an exhaustive search for an optimal Gestalt basket to future work.

\textbf{Conclusions.}
We set out to test whether the Platonic Representation Hypothesis is merely descriptive (that is, models happen to agree) or operational (that is, whether their agreement can be put to use).
We find that the PRH is useful in our tested case: our Gestalt meta-model outperforms every one of its constituents on 19 of 21 of our probed targets, strengthens monotonically as the basket grows and diversifies, is able to generalize to surveys held out of its fit, and reorganises galaxy physics into a coordinate system more physically consistent than any individual members'.
Gestalt is assembled, not trained: it costs minutes on one machine against the $\mathcal{O}(10^{4}$--$10^{5})$ A100-hour price of a fresh foundation model pre-train, ships as a $\mathcal{O}(100)$\,MB projection applied by a single matrix multiplication, and molds models the community has already trained into a common astronomical representation.
We conclude with the provocation that astronomy does not need yet another pre-trained galaxy encoder, what we need instead is to assemble, repurpose, and put to use the ample commons provided by the open source machine learning community.
We hope that Gestalt motivates further work towards that goal.


\section*{Acknowledgements}

We would like to thank Stella Biderman and EleutherAI for their compute and infrastructure support.

This research used the DeltaAI advanced computing and data resource through allocation number
PHY250286, which is supported by the National Science Foundation (award OAC 2320345) and
the State of Illinois. DeltaAI is a joint effort of the University of Illinois Urbana-Champaign and
its National Center for Supercomputing Applications. Access to DeltaAI was granted through the
Advanced Cyberinfrastructure Coordination Ecosystem: Services \& Support (ACCESS) program,
which is supported by U.S. National Science Foundation grants \#2138259, \#2138286, \#2138307,
\#2137603, and \#2138296.

\printbibliography


\clearpage

\appendix

\section{Responsible-use statement}

This work focuses on the astronomical application of existing large models, as such we do not expect it to have a negative societal impact.
We hope that the method described leads to more efficient use of the machine learning commons within the astronomy community, which will result in lower carbon emissions from redundant retraining and pre-training of models \citep{ref_lacoste2019}.

\section{Reproducibility and community}
\label{app:repro}

We welcome collaboration!

You can find the code to train your very own Gestalt here: \\\url{https://github.com/UniverseTBD/gestalt}. 

You can also join the discussion in the UniverseTBD Discord server: \\\url{https://discord.gg/SHyQxcN43u}.

\section{Full per-model tables}
\label{app:tables}

Tables~\ref{tab:cosmosweb}, \ref{tab:gz10}, and \ref{tab:galaxies}
give the complete per-member scores behind Table~\ref{tab:probes}; Table~\ref{tab:scaling-values} gives the per-cell
scaling values behind Figure~\ref{fig:scaling}.

\begin{table*}[htbp]
\centering
\scriptsize
\setlength{\tabcolsep}{4pt}
\caption{COSMOS-Web HSC and JWST linear-probe $R^2$. Best per column in bold. Values use five repeated random 40\,000/5\,000 probe splits.}
\label{tab:cosmosweb}
\resizebox{\textwidth}{!}{%
\begin{tabular}{lcccccc}
\toprule
 & $z$ (HSC) & $\log M_\star$ (HSC) & sSFR (HSC) & $z$ (JWST) & $\log M_\star$ (JWST) & sSFR (JWST) \\
\midrule
\textbf{Gestalt} & $\mathbf{0.518 \pm 0.008}$ & $\mathbf{0.604 \pm 0.012}$ & $\mathbf{0.581 \pm 0.015}$ & $\mathbf{0.658 \pm 0.014}$ & $\mathbf{0.848 \pm 0.003}$ & $\mathbf{0.482 \pm 0.007}$ \\
\textbf{Basket (concat$\to$PCA)} & \underline{$0.482 \pm 0.008$} & \underline{$0.577 \pm 0.009$} & \underline{$0.555 \pm 0.007$} & \underline{$0.595 \pm 0.013$} & \underline{$0.815 \pm 0.004$} & \underline{$0.458 \pm 0.004$} \\
\midrule
AstroPT 15M & $0.387 \pm 0.014$ & $0.525 \pm 0.010$ & $0.462 \pm 0.003$ & $0.349 \pm 0.011$ & $0.674 \pm 0.011$ & $0.293 \pm 0.009$ \\
AstroPT 95M & $0.443 \pm 0.019$ & $0.556 \pm 0.013$ & $0.502 \pm 0.007$ & $0.472 \pm 0.016$ & $0.771 \pm 0.009$ & $0.336 \pm 0.005$ \\
AstroPT 850M & $0.453 \pm 0.008$ & $0.559 \pm 0.014$ & $0.510 \pm 0.008$ & $0.531 \pm 0.013$ & $0.802 \pm 0.007$ & $0.360 \pm 0.005$ \\
CLIP 86M & $0.433 \pm 0.010$ & $0.517 \pm 0.009$ & $0.511 \pm 0.007$ & $0.473 \pm 0.010$ & $0.706 \pm 0.003$ & $0.358 \pm 0.012$ \\
CLIP 304M & $0.435 \pm 0.008$ & $0.515 \pm 0.011$ & $0.496 \pm 0.008$ & $0.533 \pm 0.010$ & $0.736 \pm 0.004$ & $0.385 \pm 0.011$ \\
ConvNeXt-V2 15M & $0.392 \pm 0.005$ & $0.490 \pm 0.010$ & $0.475 \pm 0.008$ & $0.457 \pm 0.011$ & $0.690 \pm 0.008$ & $0.359 \pm 0.012$ \\
ConvNeXt-V2 28M & $0.412 \pm 0.010$ & $0.496 \pm 0.010$ & $0.470 \pm 0.007$ & $0.455 \pm 0.018$ & $0.675 \pm 0.008$ & $0.375 \pm 0.009$ \\
ConvNeXt-V2 89M & $0.409 \pm 0.005$ & $0.501 \pm 0.009$ & $0.488 \pm 0.012$ & $0.460 \pm 0.011$ & $0.676 \pm 0.003$ & $0.386 \pm 0.007$ \\
ConvNeXt-V2 198M & $0.425 \pm 0.008$ & $0.522 \pm 0.011$ & $0.506 \pm 0.009$ & $0.498 \pm 0.013$ & $0.713 \pm 0.004$ & $0.400 \pm 0.012$ \\
I-JEPA 632M & $0.429 \pm 0.009$ & $0.528 \pm 0.008$ & $0.499 \pm 0.007$ & $0.524 \pm 0.012$ & $0.780 \pm 0.006$ & $0.403 \pm 0.010$ \\
I-JEPA 1B & $0.424 \pm 0.005$ & $0.528 \pm 0.008$ & $0.506 \pm 0.014$ & $0.555 \pm 0.014$ & $0.790 \pm 0.007$ & $0.416 \pm 0.009$ \\
LLaVA-1.5 7B & $0.431 \pm 0.011$ & $0.522 \pm 0.015$ & $0.504 \pm 0.006$ & $0.528 \pm 0.016$ & $0.752 \pm 0.006$ & $0.390 \pm 0.011$ \\
LLaVA-1.5 13B & $0.429 \pm 0.009$ & $0.518 \pm 0.010$ & $0.502 \pm 0.004$ & $0.531 \pm 0.012$ & $0.756 \pm 0.003$ & $0.394 \pm 0.013$ \\
ViT-MAE 86M & $0.343 \pm 0.011$ & $0.450 \pm 0.010$ & $0.422 \pm 0.009$ & $0.290 \pm 0.012$ & $0.581 \pm 0.010$ & $0.277 \pm 0.016$ \\
ViT-MAE 304M & $0.357 \pm 0.013$ & $0.461 \pm 0.010$ & $0.441 \pm 0.008$ & $0.324 \pm 0.013$ & $0.631 \pm 0.012$ & $0.295 \pm 0.013$ \\
ViT-MAE 632M & $0.372 \pm 0.011$ & $0.486 \pm 0.009$ & $0.470 \pm 0.006$ & $0.357 \pm 0.013$ & $0.668 \pm 0.011$ & $0.310 \pm 0.011$ \\
ViT 86M & $0.359 \pm 0.007$ & $0.455 \pm 0.015$ & $0.434 \pm 0.007$ & $0.372 \pm 0.013$ & $0.597 \pm 0.005$ & $0.306 \pm 0.013$ \\
ViT 304M & $0.369 \pm 0.007$ & $0.474 \pm 0.008$ & $0.440 \pm 0.005$ & $0.408 \pm 0.012$ & $0.638 \pm 0.007$ & $0.340 \pm 0.010$ \\
ViT 632M & $0.434 \pm 0.012$ & $0.520 \pm 0.009$ & $0.480 \pm 0.009$ & $0.477 \pm 0.008$ & $0.722 \pm 0.005$ & $0.358 \pm 0.010$ \\
V-JEPA-2 300M & $0.421 \pm 0.009$ & $0.544 \pm 0.011$ & $0.517 \pm 0.006$ & $0.525 \pm 0.011$ & $0.780 \pm 0.007$ & $0.436 \pm 0.007$ \\
V-JEPA-2 600M & $0.420 \pm 0.006$ & $0.549 \pm 0.013$ & $0.518 \pm 0.007$ & $0.521 \pm 0.013$ & $0.781 \pm 0.006$ & $0.439 \pm 0.013$ \\
V-JEPA-2 1B & $0.450 \pm 0.007$ & $0.560 \pm 0.013$ & $0.530 \pm 0.008$ & $0.565 \pm 0.012$ & $0.810 \pm 0.004$ & $0.457 \pm 0.010$ \\
\bottomrule
\end{tabular}
}
\end{table*}

\begin{table*}[htbp]
\centering
\scriptsize
\setlength{\tabcolsep}{4pt}
\caption{GZ10 linear-probe scores: macro $F_1$ for 10-way morphology and $R^2$ for photometric redshift. Values use five repeated random probe splits, with 2\,500 test and 15\,300 training objects per split.}
\label{tab:gz10}
\begin{tabular}{lcc}
\toprule
 & GZ10 class & $z_{\rm phot}$ \\
\midrule
\textbf{Gestalt} & \underline{$0.708 \pm 0.009$} & $\mathbf{0.758 \pm 0.008}$ \\
\textbf{Basket (concat$\to$PCA)} & $0.674 \pm 0.009$ & \underline{$0.729 \pm 0.010$} \\
\midrule
AstroPT 15M & $0.424 \pm 0.007$ & $0.514 \pm 0.009$ \\
AstroPT 95M & $0.471 \pm 0.007$ & $0.584 \pm 0.009$ \\
AstroPT 850M & $0.480 \pm 0.007$ & $0.630 \pm 0.017$ \\
CLIP 86M & $0.680 \pm 0.007$ & $0.664 \pm 0.009$ \\
CLIP 304M & $\mathbf{0.713 \pm 0.011}$ & $0.680 \pm 0.013$ \\
ConvNeXt-V2 15M & $0.644 \pm 0.012$ & $0.622 \pm 0.015$ \\
ConvNeXt-V2 28M & $0.677 \pm 0.008$ & $0.629 \pm 0.010$ \\
ConvNeXt-V2 89M & $0.684 \pm 0.008$ & $0.605 \pm 0.018$ \\
ConvNeXt-V2 198M & $0.681 \pm 0.002$ & $0.652 \pm 0.015$ \\
I-JEPA 632M & $0.669 \pm 0.012$ & $0.688 \pm 0.013$ \\
I-JEPA 1B & $0.681 \pm 0.009$ & $0.683 \pm 0.008$ \\
LLaVA-1.5 7B & $0.689 \pm 0.010$ & $0.608 \pm 0.019$ \\
LLaVA-1.5 13B & $0.700 \pm 0.007$ & $0.571 \pm 0.027$ \\
ViT-MAE 86M & $0.427 \pm 0.010$ & $0.500 \pm 0.016$ \\
ViT-MAE 304M & $0.468 \pm 0.004$ & $0.524 \pm 0.016$ \\
ViT-MAE 632M & $0.501 \pm 0.008$ & $0.572 \pm 0.016$ \\
ViT 86M & $0.629 \pm 0.007$ & $0.607 \pm 0.011$ \\
ViT 304M & $0.637 \pm 0.006$ & $0.637 \pm 0.011$ \\
ViT 632M & $0.586 \pm 0.008$ & $0.647 \pm 0.013$ \\
V-JEPA-2 300M & $0.693 \pm 0.007$ & $0.664 \pm 0.013$ \\
V-JEPA-2 600M & $0.672 \pm 0.012$ & $0.651 \pm 0.016$ \\
V-JEPA-2 1B & $0.698 \pm 0.005$ & $0.667 \pm 0.017$ \\
\bottomrule
\end{tabular}
\end{table*}

\begin{sidewaystable*}[p]
\centering
\scriptsize
\setlength{\tabcolsep}{4pt}
    \caption{\texttt{Smith42/galaxies} linear-probe $R^2$ on 13 targets. Values use five repeated random probe splits with a 5\,000-object test holdout per target; tight spiral uses 735 test objects from 3\,676 valid labels.}
\label{tab:galaxies}
\resizebox{\textheight}{!}{%
\begin{tabular}{lccccccccccccc}
\toprule
 & $M_g$ & $M_z$ & $g{-}r$ & $r{-}z$ & $z_{\rm phot}$ & $z_{\rm spec}$ & $\log M_\star$ & sSFR & smooth & disc & artifact & edge-on & tight spiral \\
\midrule
\textbf{Gestalt} & $\mathbf{0.650 \pm 0.012}$ & $\mathbf{0.731 \pm 0.010}$ & $\mathbf{0.863 \pm 0.007}$ & $\mathbf{0.758 \pm 0.009}$ & $\mathbf{0.779 \pm 0.005}$ & $\mathbf{0.883 \pm 0.005}$ & $\mathbf{0.669 \pm 0.012}$ & $0.367 \pm 0.032$ & $\mathbf{0.825 \pm 0.003}$ & $\mathbf{0.818 \pm 0.005}$ & $\mathbf{0.854 \pm 0.015}$ & $\mathbf{0.875 \pm 0.004}$ & $\mathbf{0.719 \pm 0.012}$ \\
\textbf{Basket (concat$\to$PCA)} & $0.613 \pm 0.010$ & $0.696 \pm 0.008$ & $0.833 \pm 0.010$ & $0.727 \pm 0.010$ & $0.747 \pm 0.008$ & $0.862 \pm 0.005$ & $0.645 \pm 0.008$ & $0.306 \pm 0.021$ & $0.790 \pm 0.008$ & $0.779 \pm 0.008$ & $0.822 \pm 0.018$ & $0.852 \pm 0.003$ & $0.662 \pm 0.031$ \\
\midrule
AstroPT 15M & $0.461 \pm 0.013$ & $0.555 \pm 0.010$ & $0.716 \pm 0.008$ & $0.599 \pm 0.008$ & $0.603 \pm 0.006$ & $0.760 \pm 0.004$ & $0.528 \pm 0.007$ & $0.247 \pm 0.030$ & $0.492 \pm 0.017$ & $0.423 \pm 0.014$ & $0.651 \pm 0.023$ & $0.464 \pm 0.005$ & $0.406 \pm 0.016$ \\
AstroPT 95M & $0.516 \pm 0.012$ & $0.613 \pm 0.010$ & $0.777 \pm 0.009$ & $0.658 \pm 0.008$ & $0.673 \pm 0.007$ & $0.807 \pm 0.003$ & $0.583 \pm 0.006$ & $0.199 \pm 0.027$ & $0.574 \pm 0.009$ & $0.514 \pm 0.011$ & $0.734 \pm 0.019$ & $0.523 \pm 0.003$ & $0.411 \pm 0.047$ \\
AstroPT 850M & $0.557 \pm 0.015$ & $0.650 \pm 0.012$ & $0.796 \pm 0.010$ & $0.691 \pm 0.008$ & $0.701 \pm 0.008$ & $0.790 \pm 0.005$ & $0.611 \pm 0.007$ & $-0.752 \pm 0.131$ & $0.636 \pm 0.009$ & $0.585 \pm 0.011$ & $0.780 \pm 0.017$ & $0.463 \pm 0.012$ & $-0.177 \pm 0.053$ \\
CLIP 86M & $0.532 \pm 0.011$ & $0.621 \pm 0.011$ & $0.766 \pm 0.007$ & $0.652 \pm 0.006$ & $0.675 \pm 0.008$ & $0.822 \pm 0.005$ & $0.590 \pm 0.018$ & $0.394 \pm 0.019$ & $0.647 \pm 0.014$ & $0.633 \pm 0.013$ & $0.739 \pm 0.013$ & $0.804 \pm 0.005$ & $0.639 \pm 0.030$ \\
CLIP 304M & $0.566 \pm 0.009$ & $0.653 \pm 0.008$ & $0.797 \pm 0.009$ & $0.679 \pm 0.009$ & $0.704 \pm 0.007$ & $0.848 \pm 0.004$ & $0.614 \pm 0.014$ & $\mathbf{0.396 \pm 0.018}$ & $0.710 \pm 0.011$ & $0.711 \pm 0.012$ & $0.765 \pm 0.010$ & $0.832 \pm 0.004$ & $0.658 \pm 0.005$ \\
ConvNeXt-V2 15M & $0.494 \pm 0.012$ & $0.579 \pm 0.011$ & $0.724 \pm 0.012$ & $0.617 \pm 0.009$ & $0.631 \pm 0.011$ & $0.791 \pm 0.004$ & $0.548 \pm 0.011$ & $0.268 \pm 0.033$ & $0.635 \pm 0.006$ & $0.621 \pm 0.007$ & $0.709 \pm 0.020$ & $0.770 \pm 0.005$ & $0.564 \pm 0.020$ \\
ConvNeXt-V2 28M & $0.522 \pm 0.008$ & $0.606 \pm 0.007$ & $0.740 \pm 0.007$ & $0.636 \pm 0.010$ & $0.657 \pm 0.007$ & $0.806 \pm 0.006$ & $0.570 \pm 0.008$ & $0.244 \pm 0.024$ & $0.656 \pm 0.010$ & $0.648 \pm 0.009$ & $0.718 \pm 0.025$ & $0.791 \pm 0.004$ & $0.566 \pm 0.010$ \\
ConvNeXt-V2 89M & $0.531 \pm 0.010$ & $0.612 \pm 0.010$ & $0.744 \pm 0.008$ & $0.637 \pm 0.009$ & $0.662 \pm 0.007$ & $0.795 \pm 0.007$ & $0.572 \pm 0.008$ & $0.130 \pm 0.035$ & $0.700 \pm 0.008$ & $0.696 \pm 0.009$ & $0.742 \pm 0.016$ & $0.810 \pm 0.004$ & $0.545 \pm 0.041$ \\
ConvNeXt-V2 198M & $0.561 \pm 0.005$ & $0.641 \pm 0.004$ & $0.778 \pm 0.010$ & $0.667 \pm 0.006$ & $0.690 \pm 0.008$ & $0.793 \pm 0.011$ & $0.597 \pm 0.010$ & $-0.251 \pm 0.065$ & $0.734 \pm 0.005$ & $0.735 \pm 0.008$ & $0.755 \pm 0.014$ & $0.798 \pm 0.007$ & $0.448 \pm 0.039$ \\
I-JEPA 632M & $0.544 \pm 0.015$ & $0.631 \pm 0.015$ & $0.766 \pm 0.007$ & $0.653 \pm 0.010$ & $0.682 \pm 0.006$ & $0.839 \pm 0.007$ & $0.593 \pm 0.012$ & $0.040 \pm 0.076$ & $0.693 \pm 0.009$ & $0.677 \pm 0.012$ & $0.754 \pm 0.016$ & $0.799 \pm 0.004$ & $0.556 \pm 0.014$ \\
I-JEPA 1B & $0.540 \pm 0.015$ & $0.628 \pm 0.013$ & $0.761 \pm 0.009$ & $0.662 \pm 0.009$ & $0.682 \pm 0.005$ & $0.830 \pm 0.006$ & $0.587 \pm 0.011$ & $0.013 \pm 0.056$ & $0.698 \pm 0.010$ & $0.676 \pm 0.014$ & $0.768 \pm 0.019$ & $0.793 \pm 0.004$ & $0.555 \pm 0.036$ \\
LLaVA-1.5 7B & $0.555 \pm 0.013$ & $0.640 \pm 0.010$ & $0.790 \pm 0.007$ & $0.678 \pm 0.004$ & $0.695 \pm 0.011$ & $0.640 \pm 0.011$ & $0.599 \pm 0.015$ & $-0.584 \pm 0.124$ & $0.716 \pm 0.007$ & $0.715 \pm 0.009$ & $0.758 \pm 0.014$ & $0.285 \pm 0.013$ & $-0.099 \pm 0.050$ \\
LLaVA-1.5 13B & $0.554 \pm 0.011$ & $0.640 \pm 0.010$ & $0.790 \pm 0.008$ & $0.678 \pm 0.005$ & $0.694 \pm 0.009$ & $0.567 \pm 0.019$ & $0.598 \pm 0.016$ & $-0.215 \pm 0.096$ & $0.727 \pm 0.007$ & $0.724 \pm 0.009$ & $0.769 \pm 0.015$ & $0.451 \pm 0.029$ & $0.273 \pm 0.020$ \\
ViT-MAE 86M & $0.448 \pm 0.010$ & $0.536 \pm 0.010$ & $0.702 \pm 0.006$ & $0.588 \pm 0.013$ & $0.597 \pm 0.011$ & $0.753 \pm 0.004$ & $0.518 \pm 0.008$ & $0.175 \pm 0.042$ & $0.552 \pm 0.008$ & $0.502 \pm 0.012$ & $0.642 \pm 0.018$ & $0.648 \pm 0.006$ & $0.384 \pm 0.034$ \\
ViT-MAE 304M & $0.475 \pm 0.009$ & $0.563 \pm 0.008$ & $0.729 \pm 0.007$ & $0.607 \pm 0.010$ & $0.632 \pm 0.010$ & $0.776 \pm 0.008$ & $0.537 \pm 0.009$ & $0.157 \pm 0.053$ & $0.592 \pm 0.010$ & $0.554 \pm 0.015$ & $0.683 \pm 0.014$ & $0.685 \pm 0.004$ & $0.393 \pm 0.015$ \\
ViT-MAE 632M & $0.494 \pm 0.009$ & $0.582 \pm 0.009$ & $0.741 \pm 0.010$ & $0.624 \pm 0.012$ & $0.650 \pm 0.006$ & $0.789 \pm 0.008$ & $0.552 \pm 0.010$ & $0.006 \pm 0.051$ & $0.640 \pm 0.009$ & $0.610 \pm 0.010$ & $0.708 \pm 0.017$ & $0.694 \pm 0.005$ & $0.409 \pm 0.014$ \\
ViT 86M & $0.516 \pm 0.006$ & $0.593 \pm 0.006$ & $0.735 \pm 0.007$ & $0.610 \pm 0.013$ & $0.656 \pm 0.010$ & $0.802 \pm 0.006$ & $0.554 \pm 0.008$ & $0.277 \pm 0.033$ & $0.669 \pm 0.007$ & $0.661 \pm 0.007$ & $0.703 \pm 0.018$ & $0.787 \pm 0.004$ & $0.587 \pm 0.021$ \\
ViT 304M & $0.546 \pm 0.012$ & $0.628 \pm 0.011$ & $0.769 \pm 0.009$ & $0.653 \pm 0.009$ & $0.689 \pm 0.008$ & $0.820 \pm 0.005$ & $0.584 \pm 0.009$ & $0.246 \pm 0.043$ & $0.709 \pm 0.012$ & $0.704 \pm 0.010$ & $0.727 \pm 0.017$ & $0.814 \pm 0.005$ & $0.619 \pm 0.010$ \\
ViT 632M & $0.563 \pm 0.009$ & $0.646 \pm 0.007$ & $0.784 \pm 0.008$ & $0.676 \pm 0.009$ & $0.700 \pm 0.008$ & $0.825 \pm 0.006$ & $0.600 \pm 0.011$ & $0.139 \pm 0.041$ & $0.698 \pm 0.010$ & $0.688 \pm 0.014$ & $0.737 \pm 0.015$ & $0.772 \pm 0.006$ & $0.461 \pm 0.037$ \\
V-JEPA-2 300M & $0.534 \pm 0.010$ & $0.620 \pm 0.007$ & $0.755 \pm 0.006$ & $0.648 \pm 0.010$ & $0.675 \pm 0.008$ & $0.837 \pm 0.006$ & $0.582 \pm 0.012$ & $0.180 \pm 0.024$ & $0.738 \pm 0.007$ & $0.723 \pm 0.008$ & $0.758 \pm 0.015$ & $0.837 \pm 0.003$ & $0.634 \pm 0.010$ \\
V-JEPA-2 600M & $0.499 \pm 0.006$ & $0.584 \pm 0.005$ & $0.715 \pm 0.007$ & $0.613 \pm 0.014$ & $0.638 \pm 0.006$ & $0.802 \pm 0.003$ & $0.550 \pm 0.010$ & $-0.024 \pm 0.089$ & $0.707 \pm 0.008$ & $0.687 \pm 0.011$ & $0.734 \pm 0.015$ & $0.782 \pm 0.006$ & $0.581 \pm 0.028$ \\
V-JEPA-2 1B & $0.537 \pm 0.012$ & $0.625 \pm 0.010$ & $0.746 \pm 0.007$ & $0.645 \pm 0.012$ & $0.665 \pm 0.006$ & $0.832 \pm 0.004$ & $0.589 \pm 0.012$ & $0.014 \pm 0.063$ & $0.720 \pm 0.007$ & $0.702 \pm 0.011$ & $0.751 \pm 0.018$ & $0.787 \pm 0.007$ & $0.594 \pm 0.031$ \\
\bottomrule
\end{tabular}
}
\end{sidewaystable*}

\begin{table*}[htbp]
\centering
\scriptsize
\setlength{\tabcolsep}{4pt}
\caption{COSMOS-Web basket-pruning $R^2$ vs basket size $k$. Each probe split is 40\,000/5\,000. The full-basket value appears at $k{=}22$ for the random series; one-per-family is undefined for $k{>}8$ because the basket contains eight model families.}
\label{tab:scaling-values}
\begin{tabular}{lllccccc}
\toprule
Modality & Property & Subset & $k{=}2$ & $k{=}4$ & $k{=}8$ & $k{=}16$ & $k{=}22$ \\
\midrule
HSC & $z_{\rm phot}$ & random & $0.420 \pm 0.029$ & $0.462 \pm 0.019$ & $0.491 \pm 0.013$ & $0.512 \pm 0.009$ & $0.519 \pm 0.008$ \\
 &  & one/family & $0.439 \pm 0.019$ & $0.468 \pm 0.014$ & $0.500 \pm 0.009$ & -- & -- \\
 & $\log M_\star$ & random & $0.521 \pm 0.026$ & $0.559 \pm 0.016$ & $0.582 \pm 0.011$ & $0.598 \pm 0.011$ & $0.604 \pm 0.012$ \\
 &  & one/family & $0.541 \pm 0.019$ & $0.563 \pm 0.014$ & $0.589 \pm 0.011$ & -- & -- \\
 & sSFR & random & $0.495 \pm 0.027$ & $0.533 \pm 0.021$ & $0.557 \pm 0.012$ & $0.575 \pm 0.011$ & $0.581 \pm 0.016$ \\
 &  & one/family & $0.511 \pm 0.017$ & $0.540 \pm 0.013$ & $0.565 \pm 0.011$ & -- & -- \\
\midrule
JWST & $z_{\rm phot}$ & random & $0.473 \pm 0.064$ & $0.562 \pm 0.044$ & $0.619 \pm 0.021$ & $0.649 \pm 0.014$ & $0.658 \pm 0.014$ \\
 &  & one/family & $0.517 \pm 0.022$ & $0.573 \pm 0.030$ & $0.633 \pm 0.014$ & -- & -- \\
 & $\log M_\star$ & random & $0.727 \pm 0.051$ & $0.785 \pm 0.031$ & $0.826 \pm 0.010$ & $0.842 \pm 0.004$ & $0.848 \pm 0.003$ \\
 &  & one/family & $0.761 \pm 0.025$ & $0.795 \pm 0.021$ & $0.833 \pm 0.005$ & -- & -- \\
 & sSFR & random & $0.372 \pm 0.032$ & $0.426 \pm 0.027$ & $0.458 \pm 0.013$ & $0.474 \pm 0.006$ & $0.482 \pm 0.006$ \\
 &  & one/family & $0.388 \pm 0.014$ & $0.430 \pm 0.018$ & $0.465 \pm 0.009$ & -- & -- \\
\bottomrule
\end{tabular}
\end{table*}

\section{Cross-survey transfer experiment}
\label{app:cross-survey}

This appendix documents the experiment summarized in Fig.~\ref{fig:native-lodo-single}: for each of the four corpora (COSMOS-Web HSC, COSMOS-Web JWST, GZ10, and \texttt{Smith42/galaxies}) we compare Gestalt representations fitted without access to the target corpus against the strongest single frozen model evaluated natively. Corpus-level scores average the target's downstream tasks (photo-$z$, $\log M_\star$, and sSFR $R^2$ for COSMOS-Web; redshift $R^2$ and morphology macro-$F_1$ for GZ10; the 13 regression-target $R^2$s for \texttt{Smith42/galaxies}) over five probe seeds.

\begin{figure}[htbp]
    \centering
    \includegraphics[width=\linewidth]
    {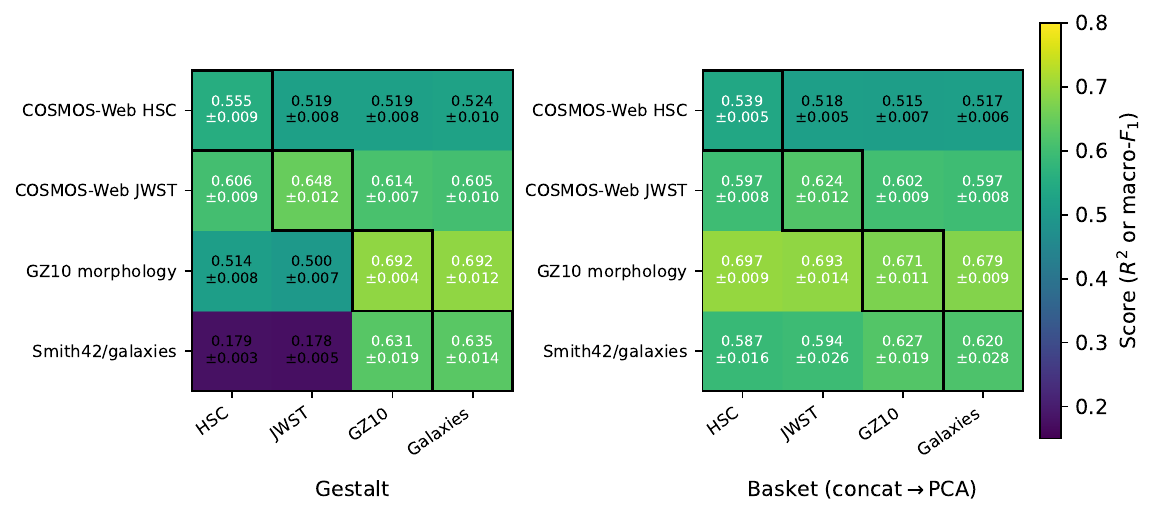}
    \caption{
        Single-source transfer across target corpora. The two coloured square matrices have four target-metric rows and four source-corpus columns; cells show mean $\pm$ standard deviation over five probe seeds, with in-domain fits outlined. The HSC, JWST, and \texttt{Smith42/galaxies} rows summarize regression $R^2$ over target tasks; GZ10 contributes its morphology macro-$F_1$ row. Each single-source fit uses the first 10\,000 rows and each target probe uses a 2\,500-object test holdout.}
    \label{fig:mixed-domain-main}
\end{figure}

\paragraph{Single-source transfer.}
For each ordered pair of distinct corpora we fit a Gestalt representation on the first 10\,000 unlabelled rows of the source corpus, freeze every per-view PCA/z-score transform and the joint projection, apply them unchanged to the full target corpus, and probe a 2\,500-object target holdout per seed. Figure~\ref{fig:mixed-domain-main} shows the complete source-by-target matrices for Gestalt and concat$\to$PCA; the orange markers in Fig.~\ref{fig:native-lodo-single} average the three off-diagonal fits per target.

\paragraph{Cross-survey Gestalt fit.}
Rather than transferring a single-domain fit, we fit one Gestalt representation on unlabelled objects pooled from the three non-target corpora; the target corpus contributes no observations to any whitening transform or to the joint projection. 
The frozen representation is applied to the target corpus and evaluated with linear probes over five seeds (blue markers in Fig.~\ref{fig:native-lodo-single}). To keep the comparison to single-source transfer conservative we test on 2\,500 objects per corpus (7\,500 in total) and find that we reach corpus-level means of 0.522 (HSC), 0.612 (JWST), 0.716 (GZ10), and 0.730 (\texttt{Smith42/galaxies}). 
This matches the mean off-diagonal single-source transfer on HSC and JWST and decisively exceeds it on GZ10 and \texttt{Smith42/galaxies}, with no target-survey imagery entering the fit.

\end{document}